\documentclass[12pt]{article}
\usepackage{graphicx}

\begin{document}
\title{The Friction of tilted Skates on Ice}
\author{J. M. J. van Leeuwen}
\maketitle
\begin{center}
Instituut-Lorentz, Universiteit Leiden,\\
Niels Bohrweg 2, 2333 CA Leiden, The Netherlands.
\end{center}

\begin{abstract}
  The friction felt by a speed skater is calculated
  as function of the velocity and tilt angle of the skate. This calculation is
  an extension of the more common theory of friction of upright skates.
  Not only in rounding a curve the skate has to be tilted, but also in
  straightforward skating small tilt angles occur. The tilt increases the friction
  substanstially and even for small tilts the increase is relevant. The increase
  of the friction with the velocity, which is very slow for the upright skate,
  becomes more pronounced for large tilts. 
\end{abstract}
keywords: solid friction, fluid mechanics, lubrication.

\section{Introduction}
Skating is an intriguing sport from the physics viewpoint, as ice seems to be the only
substance that allows skating in a remarkable range of temperatures, velocities
and skater weights. The physical problem is twofold: the ice should allow to push
oneself forward and the friction should be low enough to glide. This is achieved
by the form of the skate which has sharp edges and a thin profile, leading to a small
friction in the forward direction and a large friction in the sideways direction. It is mainly
the small friction in the forward direction that begs for an explication. 
Indeed the measured friction forces are orders of magnitude lower than the friction
between steel and another solid medium.  In spite of the
fact that skating has been around for centuries, there is still no consensus on the friction
mechanism. One school of thought \cite{faraday,weber,smit,canale} holds that the key
to skating is the structure of the  surface of ice that is ``wet'' within the temperature
range from 0 to -30 centigrade. ''Wet'' means that the surface layers of ice are
very mobile. The similar Arrhenius behaviour of the surface mobility and the friction
supports this explanation.

The more conventional school of thought \cite{lozowski, lozowski2, pomeau, vanl}
explains the low friction by the formation of a liquid layer between skate and ice.
In \cite{lozowski2} the history of the various explanations for the possibility of skating
are presented and the arguments in favour of frictional melting are summarised: skating
occurs at temperatures relatively close to the melting temperature
such that the friction provides sufficient heat to melt a thin though
macroscopic layer of ice at the contact surface of skate an ice.
The layer is also thicker than the asperities on a well polished skate.  
Frictional melting is a self-stabilising mechanism: if the friction becomes higher
the molten layer thickens and lowers the friction, while a tendency to lower friction
yields a thinner layer with higher friction. The two types of
explanation, wet surface and frictional melting, are not in conflict
with each other and could cooperate, in particular at the tip of the contact.

This paper deals with the same problem as discussed in \cite{lozowski2}: the friction
of a tilted skate. The reason to return to this problem is that in our opinion
the used  rheology needs refinement. The key quantity in deforming ice is
its hardness $p_{\rm h}$, which is the limiting pressure for elastic deformations.
Above $p_{\rm h}$ the ice deforms plastically. As skates leave a visible trail
behind, the deformation is clearly plastically. The issue is the reaction rate of
ice to an applied pressure $p$. We propose the following (Bingham) relation
\begin{equation} \label{1}
  v_{\rm ice} = \gamma (p -p_{\rm h}),
\end{equation}
where $v_{\rm ice}$ is the (downward) velocity of the ice surface and
$\gamma$ is a material constant
with the dimension [m/(Pa s)]. In Eq.~(\ref{1}) the pressure $p$ is supposed
to be larger than $p_{\rm h}$. For lower pressures the ice does not
indent plastically.
Eq.~(\ref{1}) interpolates between rheologies used in the other studies.
\cite{pomeau} assumes that ice does not indent due to pressures no
matter how high. Only melting causes a furrow, which means $\gamma=0$.
In the papers \cite{lozowski, lozowski2} it is assumed that the pressure
stays equal to the hardness, no matter how fast $v_{\rm ice}$.
This is achieved for $\gamma \rightarrow \infty$,
since then the pressure stays equal to $p_{\rm h}$.

The main impact of Eq.~(\ref{1}) is that it leads to higher pressures
on the ice, as a consequence of substantial downward velocities $v_{\rm  ice}$,
of the order of centimetres per second. The larger $\gamma$, the closer $p$
stays to the hardness. Higher pressures  
shorten the contact surface (length) and therefore lower
the friction. Thus the rheology described by Eq.~(\ref{1}) will lead
to a friction increasing with the value of $\gamma$.

Another consequence of the plastic deformation of ice in skating is
that the boundary conditions for the pressure in the water layer
differ from those used in \cite{lozowski, lozowski2, vanl} where the
pressure is set to zero at the boundary of the water layer.
We will argue that it should be equal to the hardness $p_{\rm h}$ at the
transverse boundaries of the bottom water layer.

Most studies deal with the friction of an upright skate. However the upright position
is rather rare in skating. Even in straightforward skating the skate mostly has a
small tilt angle. In curves the tilt angle may be very large i.e. substantial more than
$45^0$. There are few measurements of the friction in real skating. The only
measurements to date are of de Koning et al. \cite{schenau}, which indicate that the
upright position gives the lowest friction,
but detailed values of the friction as function of the tilt angle are lacking.
The upright position has obtained the most attention as it is easier to treat because
of the left-right symmetry and the way the ice is touched. In this
note we extend the earlier study \cite{vanl} for upright skates to tilted skates.

\begin{figure}[h]
  \begin{center}
    \includegraphics[width=0.7\linewidth]{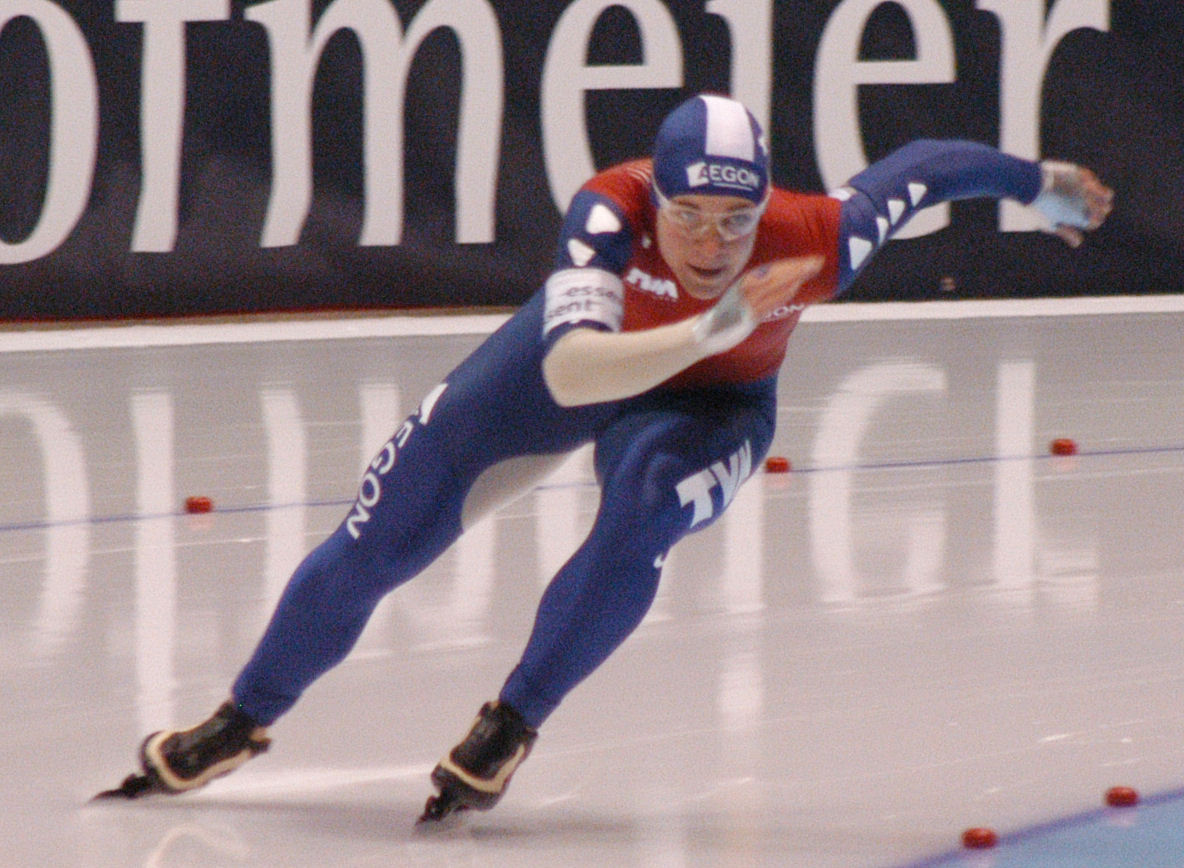}
    \vspace*{0.2cm}

    \caption{A speed skater rounding a curve. Photograph of the skater
      Paulien van Deutekom, by McSmit [CC BY-SA 3.0 Wikimedia Commons].}
\label{schul}
\end{center}
\end{figure}

The tilt angle may result from two reasons: one is that of the beginner in skating,
seeking stability from the large transverse friction. An experienced skater on the
other hand has the skate permanently in line with the legs, such that the system of body
and skate can be considered as rigid. We are interested
in the latter case.  Fig.~\ref{schul} shows a speed skater in a curve
with a rather large tilt angle. (Short trackers experience even larger tilt angles
in the curve of the track, which has a shorter radius of curvature.) The body needs
for stability a tilt angle parallel to  the resultant of the gravitational force $Mg$ 
and the centrifugal force $MV^2/R_{\rm c}$, where $M$ is the mass of the skater,
$V$ the velocity and $R_{\rm c}$ the radius of the curve.
These forces act on the center of
mass of the skater and must be compensated  by an equal and opposite force
from the ice exerting on the skate. From this equilibrium we can calculate
the tilt angle $\psi$ of the skater
\begin{equation} \label{a0}
  \tan \psi = \frac{V^2}{g R_{\rm c}}.
\end{equation} 
Although the skate is perfectly in line with the standing leg, the tilt angle of
the {\it skate} is {\it not} the same as that of the {\it body}
(defined by the line from the skate to the center-of-mass of the skater).

For the forces on the skate we need the tilt angle $\phi$ of the skate.
As the Fig.~\ref{schul} shows, $\psi$ will be somewhat
smaller than $\phi$. We will  calculate the difference. This implies that not
only the basis of the skate feels a pressure, but to a lesser extend, also the side.
Actually the way in which the skate is pushed against the ice is complicated,
as it can be varied by the muscles in the foot of the skater, without changing
the overall forces as weight and centrifugal force. The pressure can be shifted
from front to rear and from basis to side. Moreover the equilibrium to which
we alluded in Eq.~(\ref{a0}) does not need to be realised as the skater can
shift weight from one foot to the other. Here we leave all these nuances aside
and concentrate on the friction of a skate which makes a tilt angle $\phi$
with the normal and we do not consider force
components in the forward direction other than the friction between the skate
and the ice. 

We study the friction of a skate which is mainly pushed
in the ice in the direction of the tilt (see fig.~\ref{geom}).  
Lozowski et al. \cite{lozowski, lozowski2} discuss the 
the various ingredients in the melting mechanism. Here we restrict
ourselves to the two main influences: frictional melting and squeeze flow. The others,
e.g. the heat flow in the ice and skate, are of lesser importance and left out.
There are many parameters, such as: temperature, weight, speed, mass and type of
skates, which all can be varied, leading to myriad of cases. In order to avoid this,
we focus on ``standard'' skating conditions: the skater's mass $M=72 $ kg,
the velocity $V=10$ m/s, the curvature of the skate blade $R=22$ m, the width of the
blade $w=1.1$ mm and the temperature $T$ a few degrees below freezing, which we
acknowledge by the choice for the hardness of ice $p_{\rm h}=10$ MPa. 
Unfortunately the data in the literature on the hardness of ice
show a large variation \cite{pourier, new, weber}. The choice $p_{\rm h}=10$ MPa
is a compromise. For $\gamma$ we take, somewhat arbitrarily, the value
$\gamma p_{\rm h} = 2 $ mm/s in the calculations. 

Our interest is the tilt angle dependence of the friction.
After introducing the useful coordinate system, the pressure in the water layer is
derived from the hydrodynamic equations. A hydrodynamic treatment is relevant
since the water layer has a thickness of the order of a $\mu$m. Important for the
solution are the boundary conditions, for which we derive expressions in a separate
Section.
The calculation of the thickness of the water layer proceeds along similar lines as in
the case of an upright skate \cite{vanl}. The water layer at the basis and the side
have to be discussed separately, due to the different role and boundary conditions.
The paper closes with a presentation of the results
and a discussion of the main features of the solution. 

\section{The Geometry of the Indentation}

\begin{figure}[h]
  \begin{center}
    \includegraphics[width=0.7\linewidth]{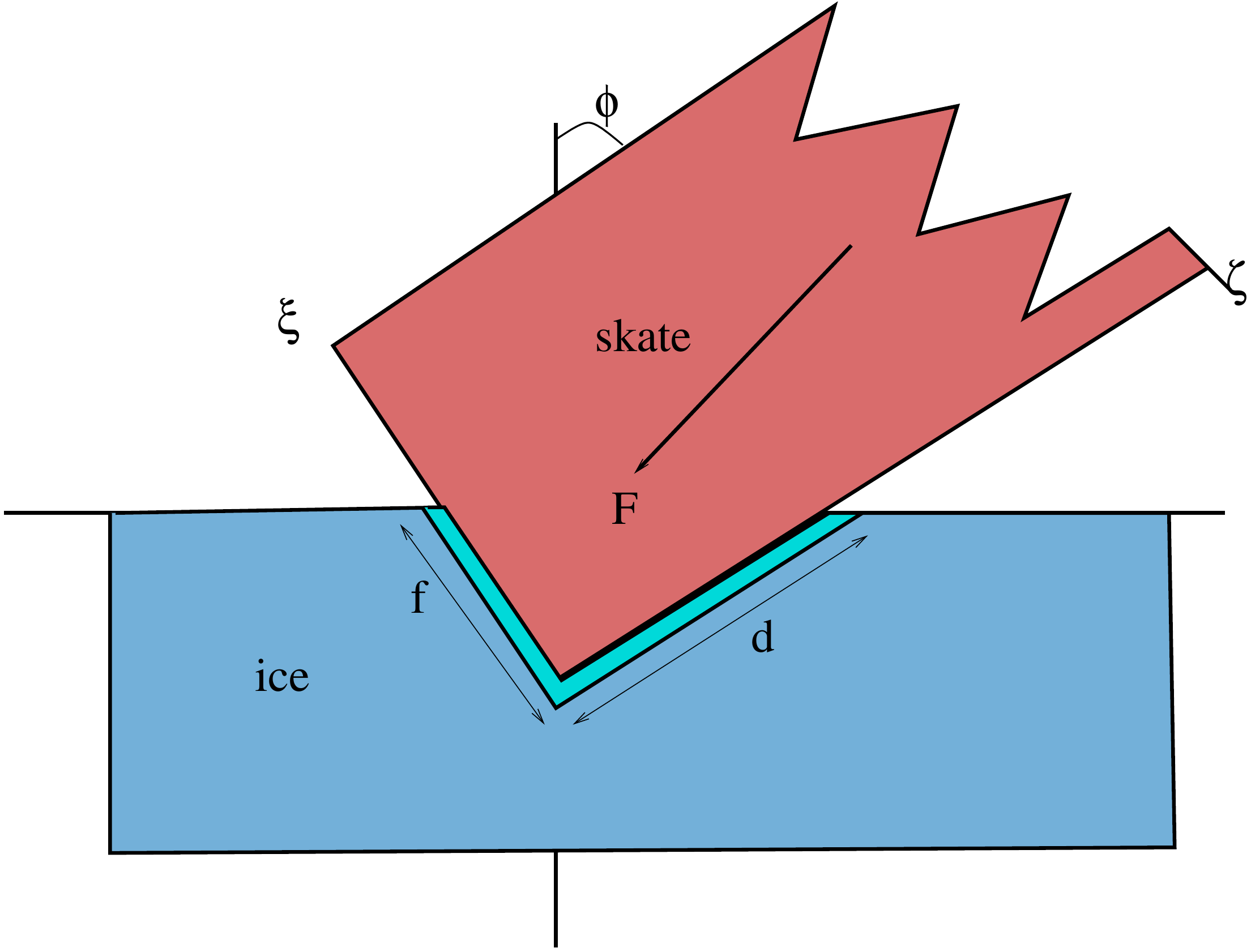}
    \vspace*{0.2cm}

   \caption{Tilted Skate indenting the Ice }
\label{geom}
\end{center}
\end{figure}

An upright skate has three contact surfaces with the ice: one at the
bottom of the skate and two at the sides. A tilted skate mostly has
two contact surfaces: one at the bottom and one at the side. Below a small
critical tilt angle $\theta_c$, one second side surface appears. We
leave out here the regime $0 < \theta < \theta_c$ as it is small and
near $\theta_c$ the friction is already close to that of the upright
skate. A picture of a transverse section of the skate is drawn in fig.~\ref{geom}.

The force drives the skate into the ice mainly in the direction of the tilt.
We separately treat the two different surfaces
between skate and ice. The bottom surface ploughs a furrow in the ice while the
skate moves parallel to the side surface.
In order to describe them it is convenient to use a coordinate system
$(x, \zeta, \xi)$ which is rotated around the $x$ axis over the tilt angle $\phi$,
see Fig.~\ref{geom}.
The $x$ axis runs along the line where the side of the skate meets the ice. The
coordinate $\zeta$ runs along the side of the skate upwardly. The $\xi$ coordinate
runs along the bottom of the skate. So for the side water layer the $\zeta$ measures
the width of the layer and the $\xi$ the thickness. For the bottom layer the $\zeta$
measures the thickness and the $\xi$ the width.

The length over which the skate makes contact with the ice the along $x$ axis
is the contact length $l$. $l$ is of the order of centimetres and thus
much smaller than the radius $R$ of the curvature of the skate.

\begin{figure}[h]
  \begin{center}
    \includegraphics[width=0.7\linewidth]{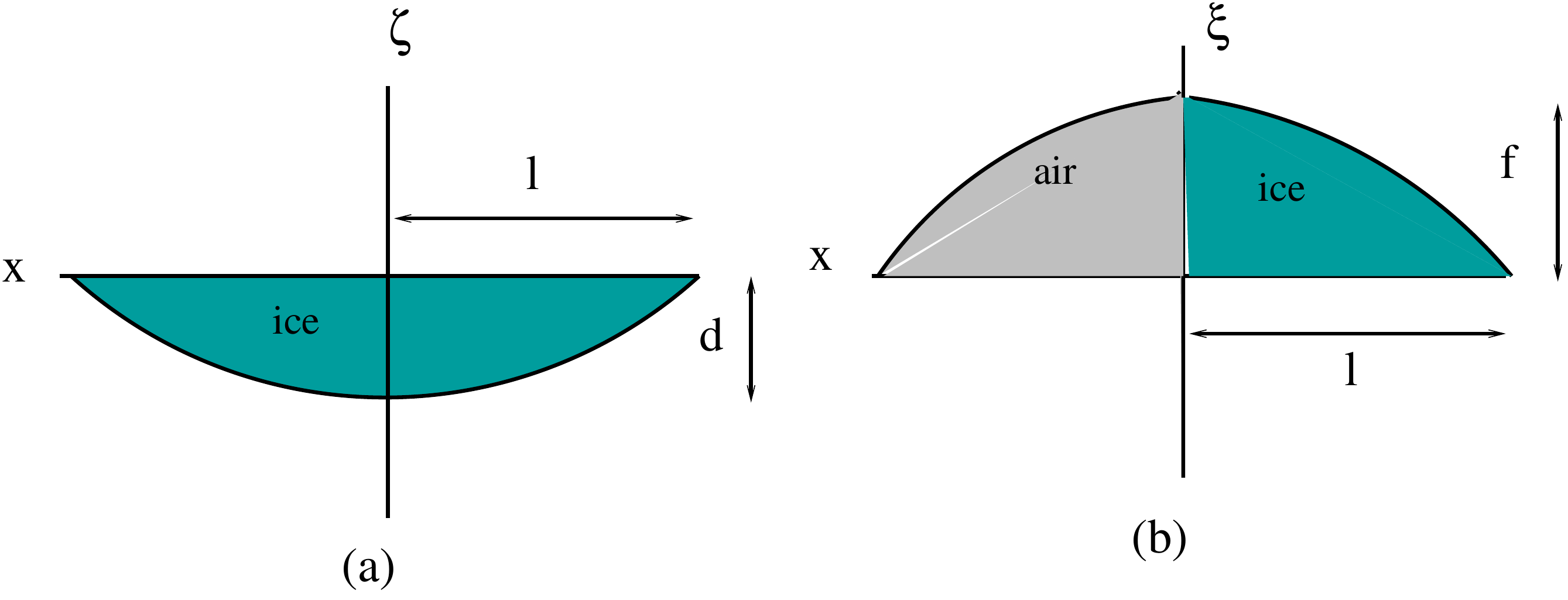}
    \vspace*{0.2cm}

  \caption{Shape of side (a) and basis (b) surface.}
\label{sides}
\end{center}
\end{figure}

The side contact surface of the skate has a simple form as it is
flat and bounded by a straight line and a circle. The straight line is
the intersection of the surface of the ice and the side of the blade
The circle is part of the edge of the skate and thus has the radius of
curvature $R$. In Fig.~\ref{sides} (a) we have drawn the side surface.
In formula the side surface is in the $(x, \zeta)$ and bounded by the
line $\zeta=0$ and the circle
\begin{equation} \label{a1}
  \zeta_e (x) = d \left(\frac{x^2}{l^2} -1 \right) \quad \quad {\rm or }
  \quad \quad x_e (\zeta)=l \sqrt{1+ \zeta/d}.
\end{equation}
Note that $\zeta_e$ is negative. 
Here $d$ is the deepest intrusion distance in the middle of the skate
and relates to the contact length $l$ as (approximating the circle by a parabola)
\begin{equation} \label{a2}
  d = \frac{l^2}{2 R} 
\end{equation} 

The bottom contact surface is more complicated as it is part of the
cylindrical bottom surface of the blade. As this cylinder has a very
large radius the bottom surface is nearly flat. On one side it is
bounded by the edge of the skate and on the other side by the
intersection of the ice surface and the bottom cylinder. In the
$(x,\xi)$ plane it is bounded by the line $\xi =0$ and the circle
\begin{equation} \label{a3}
  \xi_e (x) = f \left(1 - \frac{x^2}{l^2}\right) \quad \quad {\rm or } \quad \quad
  x_e(\xi ) = l \sqrt{1 - \xi/f},
\end{equation} 
with $f$ the distance of largest width
\begin{equation} \label{a4}
  f= \frac{d}{\tan (\phi)}. 
\end{equation}
When $f$ equals the width $w$ of the blade we reach the critical tilt
$\theta_c$. In Fig.~\ref{sides} (b) we have drawn the bottom contact surface.
We have given the coordinates of the points on the edge the subscript
$e$. We represent them either by the pair $(\zeta_e (x), \xi_e (x))$ or
reversely as $x_e (\zeta)$ or $x_e (\xi)$.

In Fig.~\ref{sides} we have given different colours to the foremost half
of the bottom surface and the other half, as in the former the skate is
separated from the ice by a melted layer and the latter also a layer
of air is in between. In the side surface there is all over a layer of
water in between.

\section{Boundary Conditions}

There is ample evidence that a skate deforms the ice plastically. The skate leaves a
visible trail behind in virgin ice and a skating rink has to be mopped up regularly in
order to improve the skating conditions. Recently Th. Boudewijn \cite{boudewijn} has
carefully measured the indentation that a skate (in the upright position) leaves behind
after it has been pushed into the ice. He finds a trough with sharp walls at the position
of the edges of the skate. So the indentation is not elastic but plastic. 
The skate ploughs a furrow in the ice, which is as deep as the skate penetrates.
Thus we ignore possible elastic deformations. 

The flow in the water layers is dominantly that of a Couette flow in the forward
($x$) direction, which is sheared on top by a skate with
velocity $V$ in the $x$ direction and sticks to the ice at the bottom.
The counter forces of the ice are mediated to the skate by water layers.
Therefore we have to know the pressure distribution in the water layers.
In Appendix \ref{layer} we summarise the standard Couette theory for the
velocity field $\bf v$ and the pressure $p$
in a layer of water of thickness $h$. There is a
force on the layer in the downward direction and due to the resulting pressure,
the water also flows in the transverse $y$ direction, with a much smaller velocity $v_y$.
The water layer is pushed down at the top with a velocity $v_{\rm sk}$ and the ice
yields with a rate $v_{\rm ice}$ at the bottom. In general the water is pushed down
at the top with a higher rate $v_{\rm sk}$ than the ice gives in at the bottom with
$v_{\rm ice}$. The difference gives the squeeze flow, which has the parabolic
Poisseuille profile.

The flow and the pressure are characterised by parameters $a,b$ and $c$.
We here only need the pressure at the top and bottom of the layer, of which
the dependence on the transverse $y$ direction is given by Eq.~(\ref{A5}) as
\begin{equation} \label{g1}
  p = \eta [-a y^2-2 b y +c],
\end{equation}
with $\eta = 1.737 \cdot 10^{-3}$ Pas. 
The parameters of the side layer are indexed as $a_s,b_s,c_s$ with the coordinate $y$
replaced by $ \zeta$ and $z$ by $\xi$. Those at the bottom are $a_b,b_b,c_b$
with coordinate $y$ replaced by $\xi$ and $z$ by $\zeta$. Thus we have to 
determine six parameters. Five of them follow from the boundary conditions
on the pressure, the parameter $a_b$ determines the rate at which
the water layer is squeezed out.

In previous calculations \cite{lozowski, vanl} the pressure was assumed to vanish
at the edges of the skate in the upright case, in the idea that it should equal the
outside air pressure, which is virtually zero as compared to the MPa pressures inside
the layer. However, measurements \cite{boudewijn} of the track left behind
by the skate, indicate sharp walls in the ice caused by the edges. So at the edge the
pressure must have been equal or larger than the hardness in order to give these
plastic deformations. Therefore we expect the pressure in the  
basis water layer to exceed the hardness $p_{\rm h}$. 

On the other hand the formation of the side layer does not involve a  (plastic)
deformation and therefore the pressure in the side water layer will be below
the hardness $p_{\rm h}$.
Continuity of the pressure then implies that the pressure at the edge,
sandwiched by the two layers, will be equal to $p_{\rm h}$.
At the point where the side layer meets open air, we assume the pressure to vanish. 

We first discuss the side layer where $y$ is replaced by $\zeta$. So the
pressure Eq.~(\ref{g1})  becomes
\begin{equation} \label{g2}
  p_s (x, \zeta) = \eta[ - a_s \zeta^2 -2 b_s \zeta + c_s]
\end{equation} 
The first observation is that $a_s=0$. It follows from the relation (\ref{B4})
between $a$ and the squeeze flow as derived in Appendix \ref{squeeze}.
Both velocities vanish: $v_{\rm sk}=0$, because the skate does not move
in the direction of the layer thickness and $v_{\rm ice}=0$, as the pressure
in the side layer will not exceed the hardness. At the edge the variable $\zeta$
in the side surface assumes the value $\zeta_e$, implying the relation
\begin{equation} \label{g3}
  \eta[-2 b_s \zeta_e + c_s] =p_h.
\end{equation}
For $\zeta = 0$ the pressure vanishes, which gives the second condition $c_s=0$.
Therefore $b_s$ has the value 
\begin{equation} \label{g4}
  b_s = -\frac{p_{\rm h}}{2 \eta \zeta_e },
\end{equation}
leading to  the following expression for the pressure in the side water layer
\begin{equation} \label{g5}
  p_s (x,\zeta) = p_{\rm h} \frac{\zeta}{\zeta_e(x)}.
\end{equation} 

For the basis layer Eq.~assumes the form
\begin{equation} \label{g6}
  p_b (x, \xi) = \eta [ - a_b \xi^2 - 2 b_b \xi +c_b]
\end{equation} 
since now $\xi$ plays the role of the transverse coordinate $y$.
For the basis layer the coordinate $\xi=0$ at the edge. This gives the condition
\begin{equation} \label{g6}
  p_{\rm h} =  \eta c_b
\end{equation}
At the other side of the basis layer, at the value $\xi=\xi_e $, we have again
the pressure $p_{\rm h}$, leading to the condition
\begin{equation} \label{g7}
  p_{\rm h} = \eta [-a_b \xi_e^2 - 2 b_b \xi_e +c_b], \quad \quad {\rm or}
  \quad \quad b_b = - a_b \xi_e/2.
\end{equation}
Combining these relations we can write the pressure Eq.~(\ref{g1}),
in the basis layer as
\begin{equation} \label{g8}
  p_b (x,\xi) = p_{\rm h} + \eta a_b \xi (\xi_e(x)-\xi)
\end{equation}

\section{Frictional Melting}

The principle of frictional melting is that the heat generated by friction melts a
layer of the ice underneath the skate. A fraction of the heat leaks away into the
skate and the other fraction melts the ice. If the temperature of the skate equals
that of the ice, both fractions are 0.5. In \cite{vanl} we have derived the 
equation (Eq.~(16)) for the layer thickness $h(x)$
\begin{equation} \label{f1}
  - \frac{\partial h}{\partial x} = \frac{k}{h} - \frac{1}{V}
  [v_{\rm sk} -v_{\rm ice}],
\end{equation}
which describes the growth of the thickness $h$ downward along the skate.
The first term on the right hand side gives the growth due to melting.
This term is characterised by the small length $k$ ($\sim 10^{-11}$m)
\begin{equation} \label{f2}
  k = \frac{\eta V}{2 \rho L_h},
\end{equation}
with $\rho$ is the density of ice $\rho = 916.8$ kg/m$^3$ and
$L_h$ the latent heat of melting $L_h=0.334 \cdot 10^6$ J/kg. The factor 2
in the denominator follows from the fact that only half of the heat is available
for melting. The second term in Eq.~(\ref{f1}) accounts for the compression
due to the rate $v_{\rm sk}$ at which the ice comes down and rate
$v_{\rm ice}$ at which ice gives in due to the pressure in the water layer.

The expression for $v_{\rm sk}$ follows from geometry of the skate. As mentioned
$v_{\rm sk} = 0$ for the side layer. The downward velocity at the bottom layer
follows from the curvature of the skate
\begin{equation} \label{f3}
  v_{\rm sk} = V \frac{\partial  \zeta (x)}{\partial x} =2 V d \frac{x}{l^2}
  = V \frac{x}{R}.
\end{equation}

\subsection{Force and Friction at the basis}

Using the imcompressibility of water we have derived in Appendix \ref{squeeze}
the relation (\ref{B4}) between the squeeze flow and $a_b$. So
Eq.~(\ref{f3}) provides the fourth equation to find pressure $p$, the parameter
$a_b$ and the two velocities $v_{\rm sk}$ and $v_{\rm ice}$.
For the bottom layer the solution for $a_b$ reads
\begin{equation} \label{f4}
  a_b = \frac{6 V x}{R [h_b^3 + 6 \gamma \eta \xi (\xi_e -\xi)]},
\end{equation}
with the associated pressure given by
\begin{equation} \label{f5}
  p_b (x, \xi) =p_{\rm h} + \frac{6 \eta V x \xi (\xi_e(x) -\xi)}
  {R [h_b^3 (x, \xi)+ 6 \gamma \eta \xi (\xi_e(x) -\xi)]}.
\end{equation} 
Hence we can make the layer equation (\ref{f1}) explicit by using Eq.~(\ref{f4})
for $a_b$
\begin{equation} \label{f6}
  -\frac{\partial h_b(x, \xi)}{\partial x} = \frac{k}{h_b (x, \xi)} -
  \frac{x h_b^3(x, \xi) }{R [h_b^3(x, \xi) + 6 \gamma \eta \xi (\xi_e(x) -\xi)]},
\end{equation}
with $\xi_e (x)$ given by Eq.~(\ref{a3}). The equation has to be integrated from
$x=x_e (\xi)$, where $h_b(x_e(\xi), \xi) = 0$, downwards to $x=0$.
So for all values of $\xi$ within $ 0 \leq \xi \leq f$ we have to integrate Eq.~(\ref{f6})
in order to find the layer thickness $h_b(x,\xi)$.
 
The bottom friction, due to the water layer is given by the integral
\begin{equation} \label{f13}
F_{\rm bw} =\eta  \int^f_0 d \xi \, \int^{x_e(\xi)}_0 dx \frac{V}{h_b (x,\xi) }.
\end{equation} 
Apart from this friction we have also the ploughing friction as a result from
making the indentation. It is given by the integral
\begin{equation} \label{f14}
F_{\rm pl} = \int^f_0 d \xi \, \int^{x_e(\xi)}_0 dx \, p_b(x, \xi) \frac{x}{R}.
\end{equation} 
The fraction $x/R$ gives the component of the force that has to be exerted 
in the forward direction. 

The normal force exerted by the basis surface on the skate equals
\begin{equation} \label{f15}
F_{\rm b} = \int^f_0 d \xi \, \int^{x_e(\xi)}_0 dx \,  p_b(x, \xi). 
\end{equation}
In these integrals the contact length enters as a trial parameter which has
to be adjusted later such that the normal force matches the weight of the
skater.
\subsection{Force and Friction on the side}

The layer equation for the water layer at the side simplifies since both
$v_{\rm sk} = 0$ and $v_{\rm ice}=0$. The former since the ice is not pushed
down at the side and the latter since the pressure stays below $p_{\rm h}$.
So the layer equation becomes for side layer $h_s$
\begin{equation} \label{h1}
  - \frac{\partial h_s}{\partial x} = \frac{k}{h_s}.
\end{equation}
The integration starts from $x_e$ at the edge, with the solution
\begin{equation} \label{h2}
  h_s(x) = [h^2_s (x_e) + 2 k (x_e - x)]^{1/2}.
\end{equation}
The initial condition $h_s(x_e)$ we find
from the requirement that the outflow of the basic layer must match in
inflow in the side layer at the edge of the skate where they meet.
The amount of water leaving the bottom layer is at an edge point ($\xi = 0$)
\begin{equation} \label{h3}
  \int^h_0 dz v_y = -b_b h_b^3 /6 = a_b  \frac{\xi_e h_b^3}{12} =
  \frac{V \xi_e x_e}{2 R}
\end{equation}
For the second equality we used Eq.~(\ref{g7}) and for the third equality 
Eq.~(\ref{f4}) with $\xi=0$. Note that the thickness $h_b$ drops out of the
relation.  The same amount of water
flows into the side layer at the same point at the edge ($\zeta = 0$)
\begin{equation} \label{h4}
  \int^h_0 dz v_y  = -b_s h_s^3/6= - \frac{p_{\rm h} h^3_s}{12 \eta \zeta_e(x)},
\end{equation} 
where we used Eq.~(\ref{g4}) for the second equality. Equating the results
(\ref{h3}) and (\ref{h4}) gives an expression for $h_s$
  \begin{equation} \label{h5}
  h^3_s(x_e) = -\frac{6 \eta V x_e  \xi_e\zeta_e}{p_{\rm h} R}.
\end{equation}
Thus the calculations of the side layer can be carried out independently
of the outcome of the basis layer, although they are connected by the
requirement of continuity of the flow at the connecting edge of the skate.

We rewrite the expression a bit with the aid of Eq.~(\ref{f2}) as
\begin{equation} \label{h6}
  h^3_s(x_e) = -12 \lambda \frac{k x_e \xi_e \zeta_e }{R}.
\end{equation}
Using the values of $\rho, L_h$ and $p_{\rm h}=10$ MPa for ice, one has
for the dimensionless combination $\lambda$
\begin{equation} \label{h7}
  \lambda = \frac{\rho L_h}{p_{\rm h}}\simeq 30.
\end{equation}
We use the result (\ref{h6}) for the solution of the layer $h_s$.
The integration of the thickness of the side water layer proceeds by fixing
a value of $\zeta$ in the region $-d \leq \zeta \leq 0$ and starting the integration
at $x_e (\zeta )$ on the edge.  Writing $\xi_e$ also as function of the
corresponding $\zeta$ via $\xi_e=-\zeta (f/d)$, the value of $h_s(\zeta)$ at the
edge becomes the initial thickness $h_0 (\zeta)$
\begin{equation} \label{h8}
  h_0^3 (\zeta) = 12 \lambda \frac{k l f  \zeta^2\sqrt{1 +\zeta /d}}{d R}.
\end{equation} 
With this value of $h_0$ the solution for the side water layer
reads more explicitly
\begin{equation} \label{h9}
  h_s(x, \zeta) = [ h_0 (\zeta)^2  + 2 k (x_e(\zeta) - x)]^{1/2}.
\end{equation}

The friction is then, for a  value of $\zeta$, given by
\begin{equation} \label{h10}
  f_{\rm fr} (\zeta) ) = \int^{x_e (\zeta)}_{-x_e (\zeta)} dx \frac{\eta V}{h_s(x, \zeta)}.
\end{equation}
The integral is elementary and reads
\begin{equation} \label{h11}
  f_{\rm fr} (\zeta) ) = 2 p_{\rm h} \lambda \left( [h_0(\zeta)^2+4 k x_e(\zeta)]^{1/2}  - 
h_0(\zeta) \right).
\end{equation}
The total friction is the integral
\begin{equation} \label{h10}
  F_{\rm sw} = \int^0_{-d} d \zeta \, f_{\rm fr} (\zeta) ),
\end{equation}
which has to be performed numerically.
The total friction is the sum of three contributions
\begin{equation} \label{h11}
F_{\rm fr} = F_{\rm bw} + F_{\rm pl} + F_{\rm sw}.
\end{equation} 

The pressure in the side layer is given by Eq.~(\ref{g5}). The force on
the side layer equals
\begin{equation} \label{h12}
  F_{\rm s} =   \int^l_0 dx  \int^0_{\zeta_e (x)} d \zeta\,  p_s (x,\zeta) =
  \int^l_0 dx  \int^0_{\zeta_e (x)} d \zeta\, p_{\rm h} \frac{\zeta}{\zeta_x} =
  p_{\rm h} \frac{d \, l}{3}
\end{equation}
The larger the tilt, the larger $d$ and the larger the force on the side of the skate. 

\section{The Force Balance}

\begin{figure}[h]
  \begin{center}
    \includegraphics[width=0.7\linewidth]{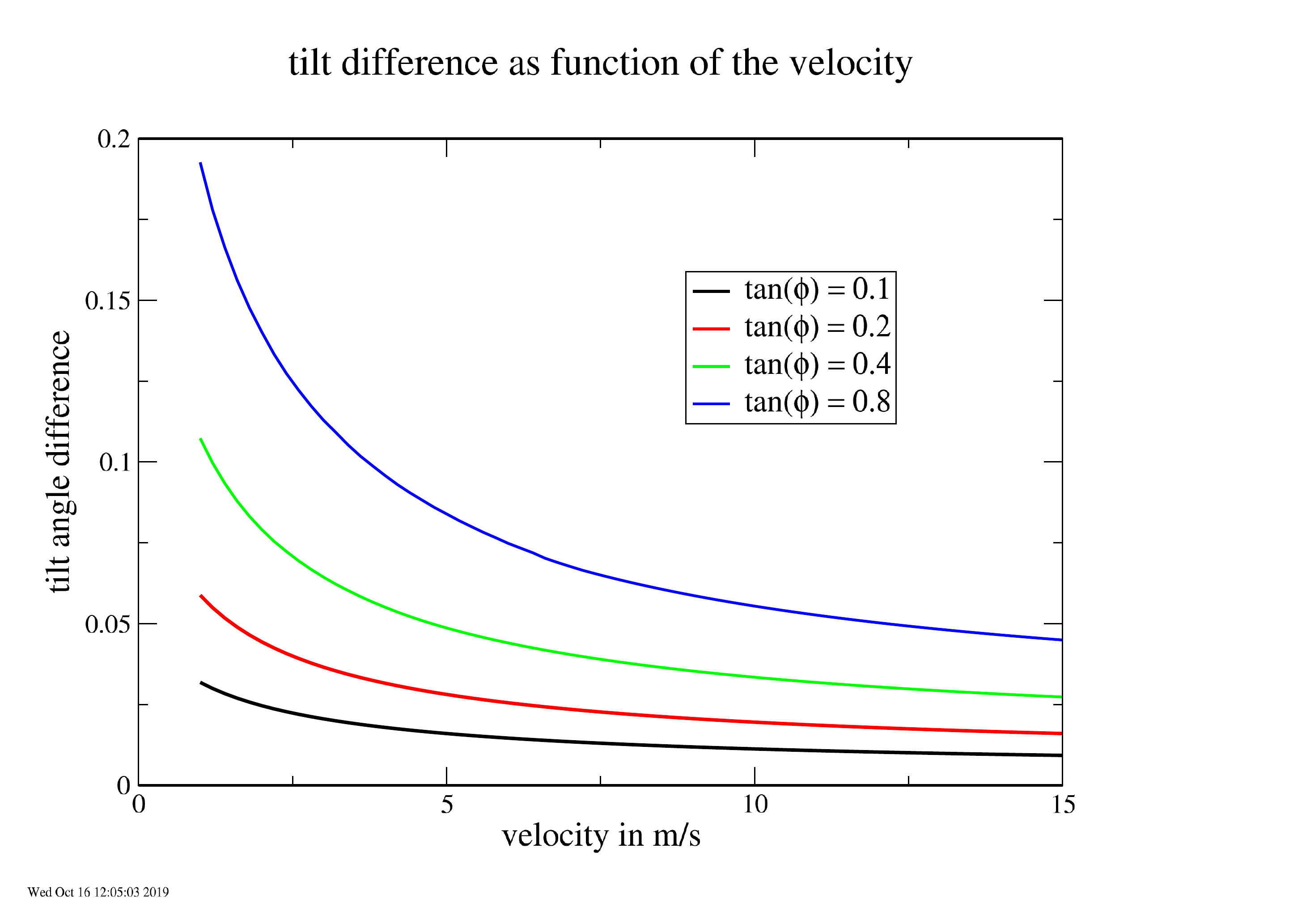}
    \vspace*{0.2cm}

  \caption{Difference $\phi-\psi$ as function of the velocity}
\label{dangle}
\end{center}
\end{figure}

The forces $F_{\rm b}$ and $F_{\rm s}$ of the ice on the skate,
given by Eqns.~(\ref{f15}) and (\ref{h12}), are normal to the surface. 
The components in the $z$ and $y$ direction are formed as the combinations
\begin{equation} \label{j1}
F_z = F_{\rm b} \cos \phi + F_{\rm s} \sin \phi, \quad \quad 
F_y = F_{\rm b} \sin  \phi - F_{\rm s}  \cos \phi.
\end{equation} 
The component in the $z$ direction balance the weight $Mg$ of the skater
and the $y$ component balances the centrifugal force
\begin{equation} \label{j2}
F_z = Mg, \quad \quad \quad F_y = \frac{MV^2}{R_{\rm c}}.
\end{equation} 
Thus we find the relation between the body inclination $\psi$ and the skate tilt $\phi$
as 
\begin{equation} \label{j3}
\tan \psi = \frac{F_y}{F_z} = \frac
{F_{\rm b} \sin  \phi - F_{\rm s}  \cos \phi}{F_{\rm b} \cos \phi + F_{\rm s} \sin \phi}
\end{equation} 
Since both $F_{\rm b}$ and $F_{\rm s}$ are positive, $\psi < \phi$, as one observes 
from Fig.~\ref{schul}. An alternative form of Eq.~(\ref{j3}) reads
\begin{equation} \label{j4}
\phi- \psi = \arctan(F_{\rm s}/F_{\rm b}),
\end{equation} 
which also follows directly from the balance of forces in the coordinate system of the
skates. In order to get an impression of these tilt differences, we have plotted, in
Fig.~\ref{dangle}, $\phi-\psi$ as function of the velocity for a number of tilt angles
$\phi$. The difference decreases slowly with the velocity and increases with the
tilt angle. For a given tilt angle $\phi$ the radius of curvature has to chosen such
that Eq.~(\ref{a0}) is fulfilled. Velocities below 1 m/s make little sense in rounding
a curve.

\section{Results}

For the calculation of the friction one has to know the thickness of the water layers.
That of the basis follows from the integration of Eq.~(\ref{f6})
in the $x$ direction for each $\xi$ in the interval $0 \leq \xi \leq f$.
If $f$ exceeds the width of the skate $w$ (as
happens for very small tilt angles), the upper limit of the $\xi$ interval has to be
replaced by $w$. The thickness of the side water layer is given by Eq.~(\ref{h9})
for the values of $\zeta$ in the interval $-d \leq \zeta \leq 0$.

\begin{figure}[h]
  \begin{center}
    \includegraphics[width=0.7\linewidth]{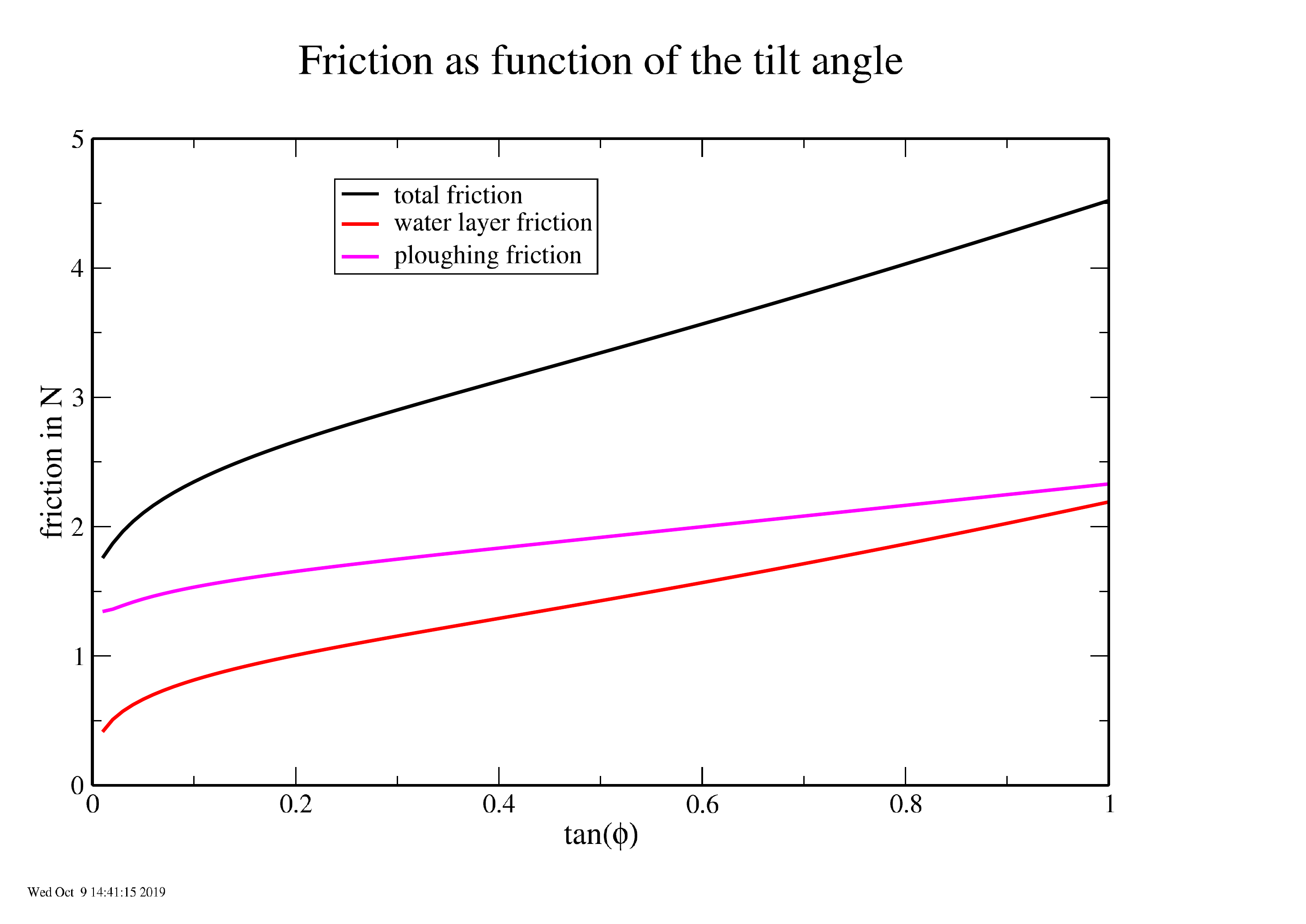}
    \vspace*{0.2cm}

  \caption{The friction contributions as function of the tilt angle}
\label{nfric}
\end{center}
\end{figure}

In general the side layer contributes a modest amount to the friction, only at
large tilt angles and high velocities it starts to count.
In Fig.~\ref{nfric} we have plotted the contributions of the friction of the water
layers and the ploughing friction, together with the total friction as function of
the tilt angle. The chosen velocity is $V=$ 10 m/s.
One observes that, while the water layer friction is rather insensitive,
the ploughing friction increases as function of the tilt angle. As a result the friction
in a curve is substantially larger friction than that of the upright skate.
This happens already for small tilt angles. One should realise that a tilt
of a few degrees easily occurs even for straightforward skating, rendering already
some 20 \% increase in the total friction.

\begin{figure}[h]
  \begin{center}
    \includegraphics[width=0.7\linewidth]{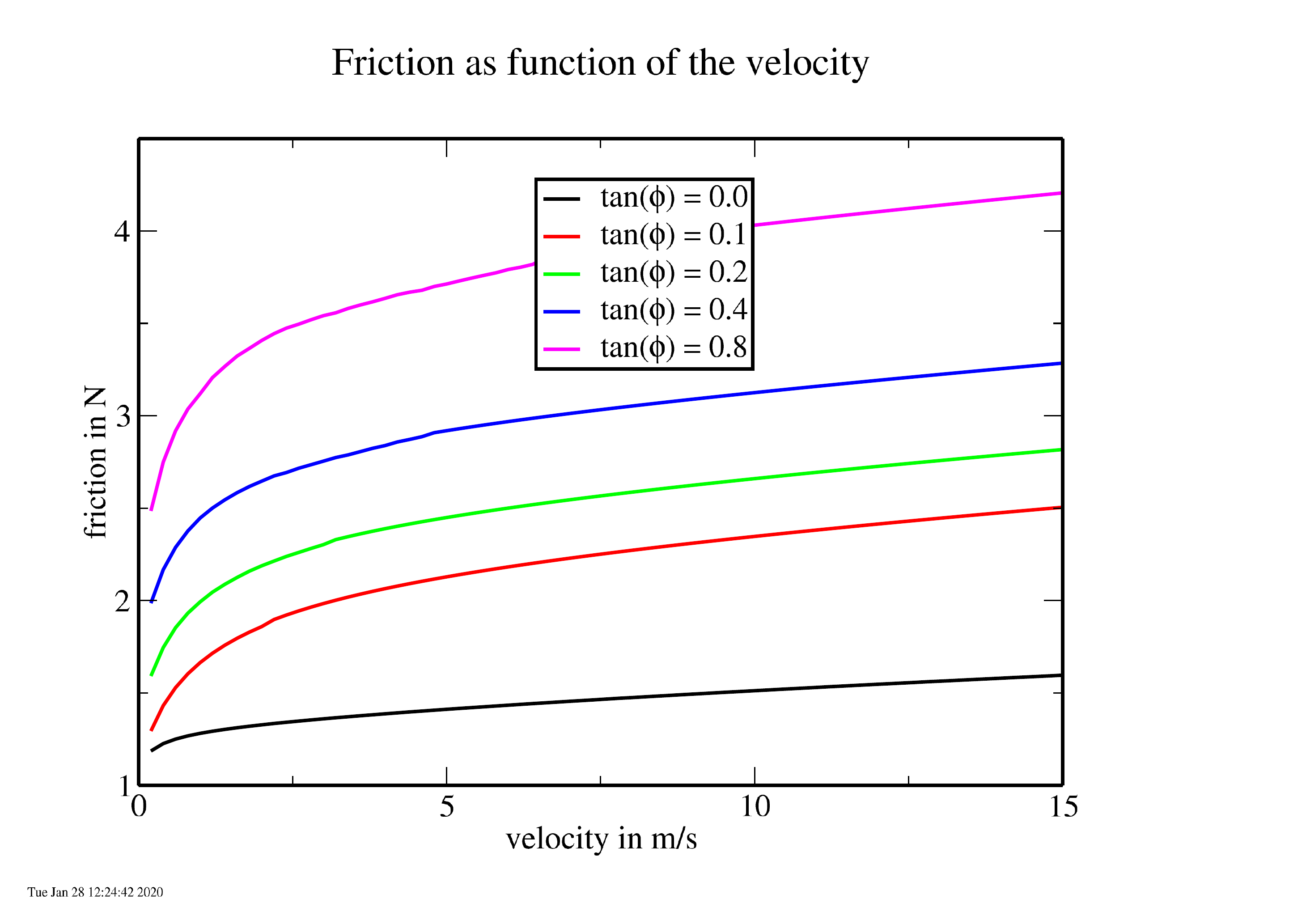}
    \vspace*{0.2cm}

  \caption{The friction  as function of the velocity for various tilt angles}
\label{curve}
\end{center}
\end{figure}

In Fig.~\ref{curve} we have plotted the friction as function of the velocity
for various tilt angles, which can be translated with Eq.~(\ref{a1})
in the radius $R_{\rm c}$ of the curve.  We see that the smaller the radius of 
the curve the larger the friction.
Apart from a region of velocities smaller than walking speed,
the increase sets in for all curvatures at higher speeds.
Note that for $\tan(\phi)=0.1$, which is a small tilt,
the friction  already rises from 1.2 at slow speeds to almost 2 for
$V$=10 m/s. This is surprising since the concomitant equilibrium curvature
of the stroke is $R_{\rm c}=100$ m 
for $V$=10 m/s, which is hardly distinguishable from a straight stroke.

By our choice of the boundary conditions on the pressure we have always
a positive contribution to the force exerted by the ice on the side of the skate.
This means that the resultant of the force on the basis and the side has a
inclination $\psi$, (slightly) different from the tilt angle $\phi$ of the skate.
The resultant points in the direction of the center of mass of the skater as
seen from the skate on the ice.

\section{Conclusion}

Using that skates trace a furrow in the ice by a plastic deformation, we 
have derived a set of equations from which the forces on the
skate can be calculated by (numerical) integration of the layer equations
for the basis and the side of the skate. We get the forces as function of the
contact length. For a giving weight of the skate one must find the
contact length by an iterative procedure 
the normal force component on the ice matching with the weight of the skater.

The problem requires boundary conditions on the pressure in the water layer,
resulting from melting due to the frictional heat. The plastic deformation 
implies that the pressure exceeds the hardness
in the basis layer and stays below the hardness $p_{\rm h}$ in the side layer.
Continuity of the pressure then gives the value $p_{\rm h}$ at the edge of the
skate. As a result we find a small force on the side of the skate.
In the equilibrium situation where the centrifugal force and the normal force
are both balanced by the generated forces of the ice on the skate, we find
the tilt angle $\psi$ of the body as function of the tilt angle $\phi$ of the skate.

Our calculation is a simplified version of the equations proposed for the upright
position \cite{lozowski, vanl} where heat generation and flows in the bulk are 
taken into account. In general these extensions are of minor influence except 
for more special circumstances as very low temperatures. We have focused on 
the influence of the tilt angle and find that the friction increases substantial with
the tilt angle. In particular, for the unavoidable small tilts occurring in skating,
the influence is large as shown in Fig.~\ref{nfric}. Note that the friction increases
with the velocity at fixed tilt angle, a feature which is stronger than the increase
found in the upright skate. The ploughing force increases with the velocity but
is tempered by following mechanism: due to the increasing pressure in the
basis water layer, the skate is lifted (aqua planing) and therefore the skate
makes a shorter contact with the ice, lowering the friction.

Comparing our approach with that the similar treatment by Lozowski
et al. \cite{lozowski2}, we observe a number of differences.
\begin{itemize}
\item Here the inclination of the skate is related to the curvature of
   the track which gives a balance between the normal and centrifugal
   force. Hence we have a force on both the bottom layer {\it and} the side
   layer, whereas \cite{lozowski2} has only a force on the bottom layer.
\item We use a different rheology, Eq.~(\ref{1}). As we pointed out in the
  Introduction, this leads generally to lower frictions. In order to give an idea we
  show in Fig.~(\ref{nfric}) also the friction for $\gamma p_{\rm h}= 2 $cm/s,
  which is ten times the value used otherwise. As one sees this change
  makes an important difference. 
\item We employ different boundary conditions for the pressure in the bottom
  water layer. We set the pressure of the bottom water layer at the boundaries in the
  transverse direction equal to the hardness, since the skate deforms the ice in an
  inelastic (plastic) way and for plastic deformation one needs pressures higher than
  the hardness. This gives also a lower friction, but less than the change in $\gamma$.
  In appendix \ref{upright} we calculate the difference for the upright skate, which
  amounts an decrease of 0.05 N  in the friction for a wide range of velocities.
\item We have left out heat leaka in the skates and ice due to unavoidable  temperature
  gradients. One could add easily these terms to the layer equation, see \cite{vanl},
  but as pointed out there, the change is not large.
\end{itemize}
The results of the calculation of \cite{lozowski2} are closer to the
measured friction in \cite{schenau} than ours, which are too low.
In general it is less worrisome to find a lower friction than a higher
friction as compared to real skating occurs, since
idealisations are made which favour gliding, such as perfectly smooth
skates and ice and since parasite processes as heat loss to the ice are omitted.
One could also interpret the difference as an indication
that our chosen $\gamma$ is too low, but we feel that the used model
is as yet too idealised to
fit parameters like $\gamma p_{\rm h}$ and $p_{\rm h}$ to the experimental results.

\appendix
\section{The hydrodynamics of the water layer}\label{layer}

We first give the shape of the flow field in a thin layer of thickness $h$ which
exist in the region between $z=0$ and $z=-h$. On top $z=0$ the velocity in the
$x$ direction equals $V$ of the skate and at the bottom $z=-h$ it vanishes (stick
boundary condition at the ice).
\begin{equation} \label{A1}
  v_x = V \left(1 + \frac{z}{h}\right),
\end{equation}
Due to the pressure water is  also squeezed out sideways 
in the $y$ direction as a Poisseuille flow
\begin{equation} \label{A2}
  v_y = -(a y+ b) z (z+h).
\end{equation}
At the top and bottom of the layer $v_y=0$. The velocity in the $z$ direction has
the profile
\begin{equation} \label{A3}
  v_z = a \left( \frac{z^3}{3}+ \frac{h z^2}{2} - \frac{h^3}{6} \right).
\end{equation}
This component is dictated by the requirement of incompressible flow
\begin{equation} \label{A4}
  \nabla \cdot {\bf v} = 0,
\end{equation}
and the condition that $v_z=0$ for $z=-h$.                                                      

The pressure which causes the flow in the $y$ direction has the form
\begin{equation} \label{A5}
  p = \eta [-a y^2-2 b y +c + a z(h+z)],
\end{equation} 
with $\eta=1.737 \cdot 10^{-3} $ Pas, the viscosity of water. One verifies
that the hydrodynamic equations
\begin{equation} \label{A6}
  \nabla p = \eta \Delta {\bf v},
\end{equation}
are fulfilled. The constants $a,b$ and $c$ have to be obtained from the boundary
conditions. 

\section{Squeeze Flow}\label{squeeze}

Consider a section $dx$ of the water layer at position $x$ with
a thickness $h(x)$ and a width $w(x)$ in the
transverse direction. The flow out of this volume is given by
\begin{equation} \label{B1}
  \frac{d\cal {V}}{dt} = dx \int^0_{h(x)} dz [v_y (w(x))- v_y (0)].
\end{equation}
Using Eq.~(\ref{A2}) we find
\begin{equation} \label{B2}
  \frac{d\cal {V}}{dt} = dx \, a\, w(x) \, h^3(x)/6.
\end{equation}
We can also compute the outflow from the difference between the rate $v_{\rm sk}$
at which the top of the volume comes down and the rate $v_{\rm ice}$ at which
the bottom of the volume goes down
\begin{equation} \label{B3}
  \frac{d\cal {V}}{dt} = dx \, w(x) [v_{\rm sk}-  v_{\rm ice}]
\end{equation}
The two expressions must agree since water is incompresible. Thus
\begin{equation} \label{B4}
  [v_{\rm sk}-  v_{\rm ice}] = a\, h^3 (x)/6,
\end{equation}
which relates $a$ to the squeeze flow.

\section{The upright skate}\label{upright}

In the upright position the skate touches the ice the
the interval $-w/2 <y < w/2$, where $w=1.1$mm is the width of the skate.
We find for the pressure, at the top or bottom of the water layer, the expression
\begin{equation} \label{U1}
p(x,y) = p_{\rm h} + \eta a(x) \left(\frac{w^2}{4}-  y^2 \right)
\end{equation}
The constant $p_{\rm h}$ raises the pressure at the edges equal to the hardness,
which was missing in the previous treatment \cite{vanl}. It was shown there that
to a good approximation we may connect $v_{\rm ice}$ with the average
\begin{equation} \label{U2}
v_{\rm ice} = \gamma \int^{w/2}_{-w/2} \frac{d w}{w} (p(x,y) - p_{\rm h}) = \gamma
\eta a(x) w^2 /6.
\end{equation} 
Combining this with the incompressibility of water, as expressed in Eq.~(\ref{f5}),
gives the relation for $a(x)$
\begin{equation} \label{U3}
a(x) = \frac{6 x V}{R(h^3(x)+ \gamma \eta w^2)}.
\end{equation} 
Then we can write the layer growth equation  explicitly as 
\begin{equation} \label{U4}
- \frac{d h(x)}{dx} = \frac{k}{h(x)} - \frac{x h^3(x)}{R(h^3(x)+ \gamma \eta w^2)}
\end{equation} 
and the pressure expression as
\begin{equation} \label{U5}
p(x)= p_{\rm h} + \frac{\eta V x w^2}{R(h^3(x)+ \gamma \eta w^2)}.
\end{equation} 
Integration of Eq.~(\ref{U4}) yields the layer thickness $h(x)$ and Eq.~(\ref{U5})
gives the pressure profile. As compared to the earlier calculation \cite{vanl}
we find a friction of 0.05 N lower over a wide range of velocities.

\end{document}